\documentclass[a4paper,11pt]{article}
\pdfoutput=1 % if your are submitting a pdflatex (i.e. if you have
\usepackage{jinstpub} % for details on the use of the package, please
\title{\boldmath CALICE SiW ECAL - Development and test of the chip-on-board PCB solution.}

\author[a]{A. Irles\note{Corresponding author.} on behalf of the CALICE collaboration}
\affiliation[a]{Universit\'e Paris-Saclay, CNRS/IN2P3, IJCLab, 91405 Orsay, France}

\emailAdd{irles@lal.in2p3.fr}
\abstract{
    The highly granular silicon-tungsten electromagnetic calorimeter (SiW ECAL)
    is the reference design of the ECAL for the International Large Detector (ILD) concept,
    one of the two detector concepts proposed for the future International Linear Collider (ILC).
    Prototypes for this type of detector are developed within the CALICE Collaboration.
    This contribution reports on the development of an ultra-thin PCB
    called Chip-on-Board (COB) that is equipped with wirebonded ASICs and pixelated silicon wafers.
    This set will form the basic unit of detection of the SiW ECAL.
    It will favour the design of an ultra compact calorimeter since these boards
    feature a thickness of 1.2\,mm to be compared with the 3\,mm of the default solution variant using ASICs
    in BGA packaging.
    We report also on the first results obtained with the COB boards on a beam test at DESY.
}


\proceeding{Calorimetry for the High Energy Frontier 2019\\
  November 25-29, 2019\\
  Fukuoka, Japan}

\begin{document}
\maketitle
\flushbottom

\section{Introduction}

The next large accelerator based particle physics experiment
will, most possibly, be an  $e^{+}e^{-}$ collider at a relatively high energy. 
Projects of different natures are currently under discussion. 
One particular example is the International Linear Collider (ILC) which has produced a technical
design report (TDR) in 2013~\cite{Behnke:2013xla}. This project offers a wide high
precission physics program
based on collisions of polarized electron and positron beam
at a nominal centre-of-mass energy of 250 GeV and with potential runs at energies spanning between 91 GeV - 1 TeV.
Therefore, the ILC will be Higgs boson and $f\bar{f}$ factory. 

To accomplish the ambitious physics program of the ILC, two multipurpose detectors 
have been proposed: the International Large Detector (ILD) and the Silicon Detector (SiD)\cite{Behnke:2013lya}. 
Both detectors are optimized to use the Particle Flow (PF) techniques \cite{Brient:2002gh,Morgunov:2004ed}.
These techniques require maximizing the information provided in each collision in order to 
fully reconstruct and separate all particles generated. This implies
the construction of detectors with high granularity and featuring minimum dead material.
This is particularly challenging for the calorimetric systems, which also are required to be very compact.
The R\&D of highly granular calorimeters for future linear colliders is conducted within the CALICE collaboration \cite{calice}.
For further information about PF and CALICE R\&D we refer the reader to reference \cite{Sefkow:2015hna} and references therein. 

In this document, we discuss some of the developments of
the technological prototype of the silicon tungsten electromagnetic calorimeter~\cite{KAWAGOE2020162969}, SiW ECAL.
This calorimeter is the reference design for the electromagnetic calorimeter of the ILD.
It will
be placed inside a magnetic coil that will provide 3.5-4 T. 
The baseline design consists in a detector of 20-24 radiation lengths ($X_{0}$) 
integrated in a volume of $\sim 20$ cm. In this volume, the SiW ECAL
has to contain the active material (silicon, Si) and the absorber (tungsten, W).
The SiW ECAL will feature $\sim 10^8$ channels, each one reading out one of the
5$\times$5 mm$^{2}$ squared cells in which the silicon sensors will be segmented in.
Due to these requirements, the SiW ECAL design foresees that the very-front-end (VFE) electronics will be
embedded in the modules together with the silicon sensors, the PCB and the tungsten plates. Each of these detector layers will
have, in average, 10000 channels. The data acquisition will be based on self-triggering and zero suppression mode for all channels independently.
The strict space constrains leaves no space to any active cooling system between modules and therefore all heat has to be dissipated through the
structure. Therefore, the overall power consumption has to be reduced to the minimum. For that, the SiW ECAL
will exploit the special bunch structure
foreseen for the ILC: the $e^{+}e^{-}$ bunch trains will arrive within
acquisition windows of $\sim$ 1-2 ms width separated by $\sim$ 200 ms. During the idle time, the bias currents of the electronics will be switched off.
This technique is usually denominated power pulsing.
In addition, only a volume of about $6\times 18 \times 0.2$ cm$^3$ is available for the digital readout 
and the power supply of the individual detector layers.

In this document we focus on the development
of a new 1.2 mm thick 11-layer PCB with wirebonded ASICs denominated COB (chip-on-board).
We also presented in this conference a newly developed front-end system based on the SL-Board readout and interface card.
This is described in \cite{Breton:2020xel}. All COB boards
were readout using this new DAQ in a performance beam test at DESY,
as discussed in section \ref{sec:TB}.
The performance of these objects at beam test at DESY is discussed in Section \ref{sec:TB}. 

\section{Active Signal Unit (ASU)}
\label{sec:COB}

The entity
of sensors, thin PCB (printed circuit boards) and ASICs (application-specific integrated circuits) is called Active Signal Units or ASU.
Each individual ASU has a lateral dimension of 18x18 cm$^{2}$. Four silicon wafers
are glued onto it. The ASU is equipped with 16 ASIC for the read out and features 1024 square pad sensors ({\it p} on {\it n}-bulk type) of 5.5x5.5 mm.
The ASICs are the version 2a of the SKIROC \cite{Callier:2011zz,Suehara:2018mqk}
(Silicon pin Kalorimeter Integrated ReadOut Chip) ASIC which has been designed for the readout of the Silicon PIN diodes.
This ASIC, as the version 2, works in self-trigger mode but still without automatic zero suppression.
Each ASU can hold up to 4 sensors of 90$\times$90 mm$^2$, each subdivided in 256 pads.
Each of the sensor pads is connected to the ASU pads through a dot of conductive glue.
The sensors are glued to the PCBs as explained in \cite{KAWAGOE2020162969}.
The bias voltage needed for the sensor depletion is provided through a conductive foil of copper and kapton
glued to the back of the sensor and connected to the high voltage through the interface card or the SL-Board. 
All previous published results of the technological prototype of the SiW ECAL \cite{KAWAGOE2020162969,Amjad:2014tha} have been obtained
with ASUs equipped with ASICs in different plastic/ceramic packagings. 
This generation of ASUs is denominated as FEV. 
So far, the most compact versions
were the FEV10-13 ASUs \cite{KAWAGOE2020162969,Poschl:2019ppa,yumiura}, equipped with 16 BGA packaged ASICs.
These boards have a thickness of 3.2-3.5 mm including components and connectors.

\subsection{Ultra thin ASU based on the chip-on-board (COB)}
\label{sec:COB}

\begin{figure}[!t]
  \centering
  \begin{tabular}{l}
\includegraphics[width=3.0in]{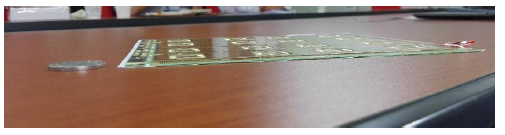}\\
\includegraphics[width=3.0in]{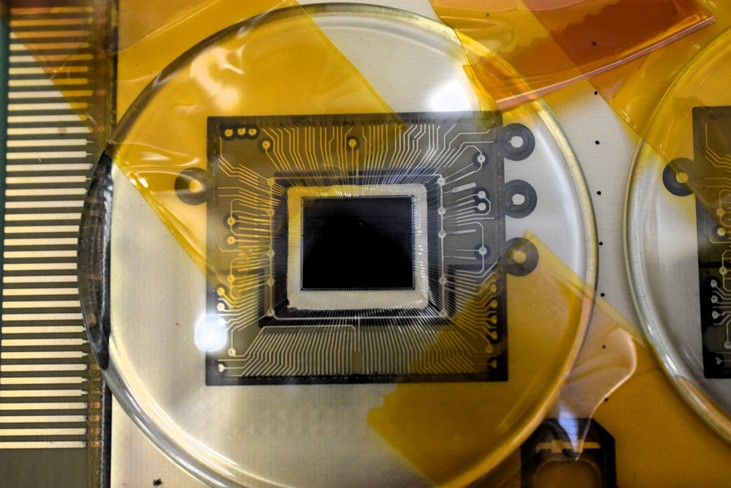}
\end{tabular}
\caption{Upper figure: one of the COB ASUs on a table, before ASIC wire bonding. This board is the result of a collaboration between the LAL and Omega institutes from France (CNRS-IN2P3) with the Sungkyunkwan University and the EOS Corporation from South Korea. Lower figure: detail of one Skiroc 2a wirebonded in one of the COBs. To protect the ASIC but keeping visual access to the wires, a watch glass has been added in top of it. For future productions it is foreseen that the ASICs will be protected with a synthetic resin.}
\label{cob}
\end{figure}

An alternative concept for the ASU has been proposed and produced. It consists on a 
ultra thin 11-layer alternative PCB design in which the ASICs
in semiconductor packaging are 
directly glued on board of the PCB in dedicated cavities and wirebonded to the PCB.
This concept of ASU is denominated chip-on-board ASU or COB. It has been designed to be compatible with
the batch of ASUs that were produced and operative at the time of the design (FEV10-12), {\it i.e.} featuring the
same dimensions, number and location of pads, connectors, ASICs, same high voltage supply scheme etc.
For the beam test described in Section \ref{sec:TB}, two COB boards were equipped with 
a $90\times 90$ mm$^{2}$ Si sensor of 500$\mu$m thickness. As for the rest of FEVs
the sensors are glued to the PCBs as explained in \cite{KAWAGOE2020162969}. 
Due to the fragility of the sensor, an excellent planarity of PCB is required in order to not damage the
sensor through mechanical stress once it is glued. This is satisfied by all COBs produced
in this last batch, showing a deviation from the planarity equal or lower than 0.5 mm.
Two photographs showing the COB are shown in Fig. \ref{cob}.
This made possible the gluing of the silicon sensors to the back of the COB
by an automatized gluing robot.

It is important to notice that due to the tight specifications on the total thickness of the PCB,
the COB ASU does not allow any space for extra components such as decoupling capacitances for the
ASIC power supplies, in contrast with the FEV generation that has up to 4 decoupling capacitances of few 100 $\mu$F per ASIC
in order to reduce disturbances in the power supplies that create spurious signals that may compromise 
the data acquisition. These signals are observed as consecutive triggers bunchs that may involve 
many channels and may fill out the 15 analogue memory cells of the ASIC quicker than the real signals. They are usually denominated 
as {\it retriggers} and they are described in detail in \cite{Amjad:2014tha}.
If they are not saturating the
memory of the ASICs,
these {\it retriggers} can easily be filter-out by studying the time distribution of the triggers.

\section{Performance in an electron beam test}
\label{sec:TB}

For the beam test conducted at DESY in summer 2019, up to 9 ASUs were equipped and tested. 
This was also the first beam test in which the new ultra compact front-end solution based on the SL-Board system
was used.
The nine ASUs were hosted in the same mechanical structure which was designed
to provide maximal flexibility to adapt to the different dimensions of the different modules and
also to the different front-ends. The mechanical structure, made in aluminum and plastic,
had two different patch panels to host the new and old front-end solutions.

We tested two COBs. Both of them were only partially equipped with
one silicon sensor of 500 $\mu$m of thickness $90\times90$ mm$^{2}$ of surface. In addition,
one of the COB was equipped with four extra decoupling capacitances (120 $\mu$F each) between the analogue power supply
layer of the board and the ground.
We placed these decoupling capacitances near the connector pads, since the board has no space for them near the ASICs
{\it i.e.} effectively increasing the total thickness of the PCB. The purpose of adding these capacitances was to compare the performance
of both COBs in order to establish what are the minimal requirements of a new design iteration.
The rest of the ASU were of the FEV generation: two FEV12 and five FEV13. The FEV12 were each one equipped with a silicon sensor of 500 $\mu$m of thickness $90\times90$ mm$^{2}$ as the COB.
Four of the FEV13 were each one equipped four silicon sensors of 650 $\mu$m of thickness.
The last FEV13 was equipped with four silicon sensors of 320 $\mu$m of thickness.

\begin{figure}[!ht]
  \centering
  \begin{tabular}{c}
	\includegraphics[width=3.0in]{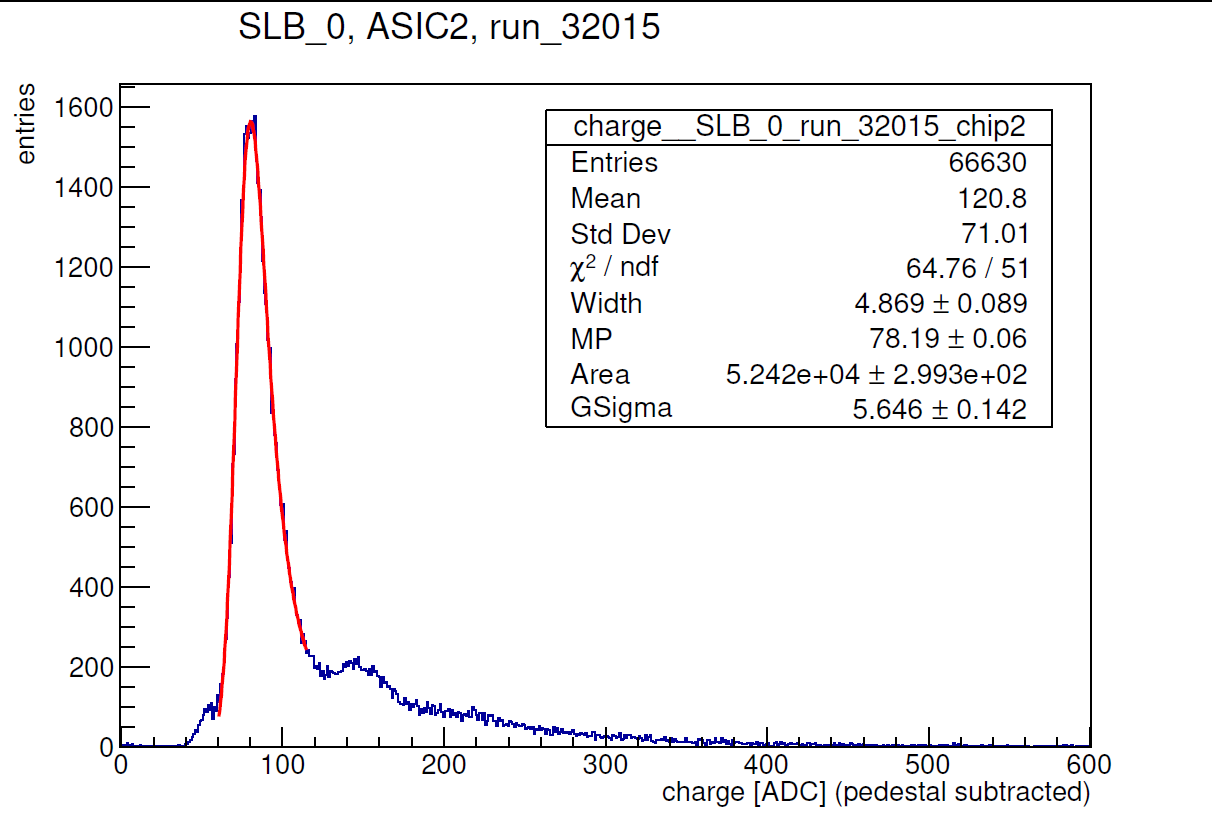} \\
	\includegraphics[width=3.0in]{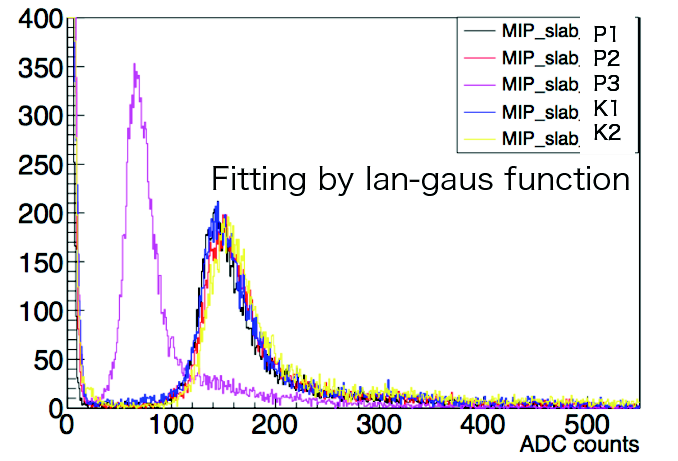}\\
	\includegraphics[width=3.0in]{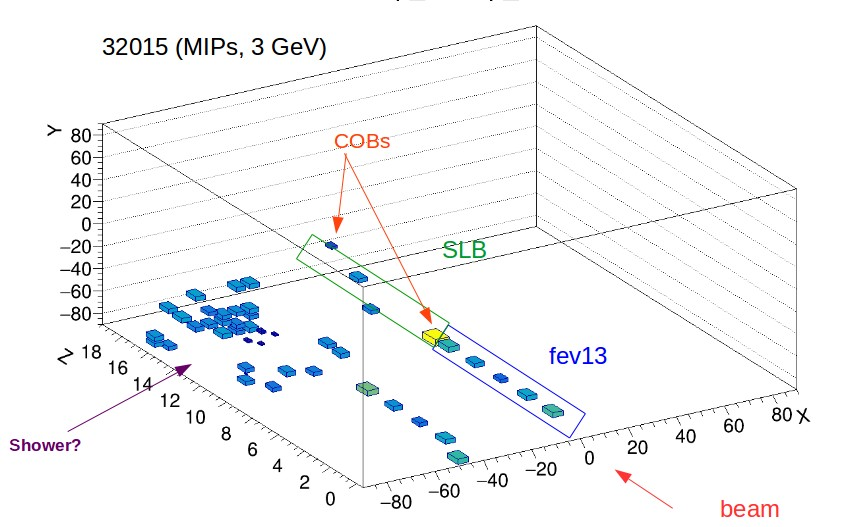}
\end{tabular}
  \caption{First plot: MIP distribution for all channels readout by one ASIC of the COB boards with extra decoupling capacitances.
    We observe in this plot a bump around the expected value of 2 MIPs:
    it is due to events in which two electrons cross together the detector.
    Second plot: equivalent distributions obtained during the same run for all the FEV13s modules.
    In this case, the small peak in 2 MIPs is not observed, probably due to a larger
    spread of the calibration values between cells.
    Last plot: event display, again for the same run, showing the signal of all modules for the same event. We see that two electrons
    separated by $\sim$50 mm interacted with the detector at the same time.
    One of them even started to shower in the middle of the detector.
    This event display was obtained before performing any offline module alignment. }
\label{mips}
\end{figure}

The beamline at DESY provides continuous electron beams in the energy range of 1 to 6 GeV with rates from
a few hundreds of Hz to a few kHz with a maximum of $\sim$3  kHz  for  2-3  GeV.  
The full program took two weeks and most of it was dedicated to the study of the MIP response of the different
modules. This can be seen in the first two plots in Fig. \ref{mips}. There we show the MIP spectrum for
one of the COBs and the FEV13s. In both cases, the same noise filtering, pedestal correction and fit distribution procedures 
as explained in \cite{KAWAGOE2020162969} are applied. We observed that the COB in which we had added
extra decoupling capacitances feature a comparable level of {\it retriggers} than the
ASUs based on the BGA packaged ASICs, making possible the use of all
the 15 analogue memory cells of the ASIC.
For the COB without decoupling capacitances, only the last 13 of these memory cells were usable.
It is important to remark
that FEV boards based in BGA packaged ASICs have more than ten times more decoupling capacitances for noise filtering.
Detailed studies on this issue are being conducted and will be the main topic of a future beam test.

\section{Conclusions and prospects.}

In this document we have summarized the status and performance in beam
test of SiW ECAL prototype of CALICE. 
In particlar, we report on the development and test of the ultra thin PCB 
denominated COB. This board is the result of a long R\&D process but this is the first
time to be operated in a beam test. The performance of this board is very promising
and it seems to be competitive with the other type of PCBs, although more detailed
test in a beam facility will be conducted to study it in better detail.
A new beam test is planned in March 2020 also at DESY. For this beam test, a calorimeter
of up to 15 modules will be tested using the new DAQ.
Among these 15 modules we foresee to have 2-3 being based on the COB solution.

\acknowledgments

This project has received funding from the European Union{\textquotesingle}s Horizon 2020 Research and Innovation program under Grant Agreement no. 654168.
This work was supported by the P2IO LabEx (ANR-10-LABX-0038), excellence project HIGHTEC,
in the framework {\textquotesingle}Investissements d{\textquotesingle}Avenir{\textquotesingle}
(ANR-11-IDEX-0003-01) managed by the French National Research Agency (ANR).
The research leading to these results has received funding from the People Programme (Marie
Curie Actions) of the European Union{\textquotesingle}s Seventh Framework Programme (FP7/2007-2013)
under REA grant agreement, PCOFUND-GA-2013-609102, through the PRESTIGE
programme coordinated by Campus France.
The measurements leading to these results have been performed at the Test Beam Facility at DESY Hamburg (Germany), a member of the Helmholtz Association (HGF).

\bibliographystyle{JHEP}
\bibliography{../references}

\end{document}